**Title:** A model for integrating the effects of multiple stressors on marine ecosystems

**Authors:** Richard M. Bailey, Jesse M.A. van der Grient

School of Geography and the Environment, University of Oxford, Dyson Perrins Building, South Park Road, Oxford, OX1 3QJ, UK

**Corresponding author:** Richard Bailey; Tel: +44 (0)1865 285530; Email: richard.bailey@ouce.ox.ac.uk

Abstract

While much has been learnt about the impacts of specific stressors on individual marine organisms, considerable debate exists over the nature and impact of multiple simultaneous stressors on both individual species and marine ecosystems. We describe a modelling tool (OSIRIS) for integrating the effects of multiple simultaneous stressors. The model is relatively computationally light, and demonstrated using a coarse-grained, non-spatial and simplified representation of a temperate marine ecosystem. This version is capable of reproducing a wide range of dynamic responses. Results indicate the degree to which interactions are synergistic is crucial in determining sensitivity to forcing, particularly for the higher trophic levels, which can respond non-linearly to stronger forcing. Stronger synergistic interactions sensitize the system to variability in forcing, and combinations of stronger forcing, noise and synergies between effects are particularly potent. This work also underlines the significant potential risk incurred in treating stressors on ecosystems as individual and additive.

Highlights

- A new flexible modelling framework to incorporate effects of multiple stressors on marine ecosystems is developed
- Importance of synergies between forcings in determining the future trajectory of marine ecosystems is demonstrated
- Stronger synergistic interactions sensitize the system to variability in forcing
- This work underlines the significant potential risk incurred in treating stressor effects on ecosystems as individual and additive

## 1.0 Introduction

In the last three decades, climate change has been acknowledged as one of the greatest threats to the natural world, and to human wellbeing. Anthropogenic greenhouse gas emissions, and $CO_2$ specifically, are widely recognized as the biggest long-term threat to functional oceans (Rogers and Laffoley 2013), due to the suite of associated impacts. Ocean acidity is being driven up, influencing large swathes of ocean ecosystems in a range of ways (Suggett and others 2012; Kroeker and others 2013). The ocean has absorbed over 90% of the excess heat from global warming, with consequences for organisms that are adapted to specific temperature ranges both in terms of latitudinal range as well as depth ranges (Stocker and others 2013). Oxygen content is in decline overall, with a dramatic increase in extreme (hypoxic) events (Stramma and others 2010). Other additional and often localized stressors include habitat destruction, overfishing, and pollution of various kinds.

While there is a mounting body of evidence on the impacts of these single stressors on single species, considerable debate exists over the impact of multiple simultaneous stressors in marine environments, and the nature of their combined effects on both individual species and on the vulnerability of wider ecosystem functions (Crain and others 2008; Piggott and others 2015; Côte and others 2016). It is likely that stress on marine ecosystems will increase over the coming decades, with far-reaching implications for conservation and resource management (Worm and others 2006; Halpern and others 2008, 2015). The potential novelty of future stressors (in type, extent and combination) means it is not necessarily possible to 'look up' past example responses in any given context. In these cases, modelling can provide valuable insights (Polovina 1984; Fulton 2004; Christensen and others 2008). Here we describe a modelling tool which incorporates the effects of multiple simultaneous stresses on marine ecosystems, and particularly those stresses associated with human activity: impacts from global level climate change associated with greenhouse gas emissions, which is driving changes in water temperature, pH, and dissolved oxygen content (Pörtner and others 2014); and more localised factors such as pollution, habitat modification, and fishing pressure. Changes (typically increasing) in these stressors are well-documented (Vitousek and others 1997), as are their effects on both individual species groups (Rodolfo-Metalpa and others 2011; Parker and others 2013) and on whole ecosystems (Hoegh-Guldberg and Bruno 2010).

While examples exist of modelling evidence providing relatively fine-scale predictions of the effects of individual stressors on whole ecosystems (e.g. pH, Marshall and others 2017), gaps exist in our ability to predict the combined effects of multiple stressors more generally, both at global scale (using Earth-system models) and local scale (using process-based trophic models). Development of multiple-stressor models is a difficult proposition, largely because of uncertainty in the basic science needed to support such an effort. This uncertainty includes the direct effects of individual stresses on individual species and also the nature of interaction effects between stressors, be they additive (total effect equal to the sum of individual effects), antagonistic (total effect less than the sum of individual effects) or synergistic (total effect greater than the sum of individual effects). Meta-analyses of empirical observations paint a complicated picture, in which the full range of interaction types are observed, and where the nature of specific interactions may depend on the presence of others (Crain and others 2008; Piggott and others 2015; Côte and others 2016). A modelling framework is needed with which to make first-order estimations of the possible effects of multiple stressors, and this framework must also facilitate the exploration of both structural uncertainty (elements present in the model and their interactions) and parameter uncertainty (random plus systematic uncertainty in parameter values).

Our approach is to build a relatively simple coarse-grained (and non-spatial in the first instance) 'whole system' model, that is sufficiently computationally light to allow parameter uncertainty analysis through large-scale statistical re-sampling, and adaptable enough to allow for efficient model-building and relatively easy re-structuring. In this paper, our aim is to outline the nature of the model, demonstrate model behaviour using both abstract cases and a simple temperate marine ecosystem representation, present examples of the kinds of output the model produces and the insights that follow.

## 2.0 Model outline

The OSIRIS (Ocean System Interactions, Risks, Instabilities and Synergies) model represents ecological systems as a network of interconnected nodes. It is a coupled-ODE (ordinary differential equation) model, and the basic outputs are time-series of the condition of each node. Nodes are loosely defined as representations of groups of elements with shared characteristics, and are either biotic nodes, which are biological populations associated with species groups, or abiotic nodes, which are typically nutrient/chemical concentrations. This loose definition is intended to allow flexibility. For example, biotic nodes can be aggregated age cohorts within individual species populations (with links to other similar nodes appropriate for their life stages). Alternatively, nodes could represent spatially segregated individuals or sub-populations within species. At a coarser level, nodes can be aggregations of large numbers of functionally similar species (and a hierarchical structure of nodes is of course allowed within this framework). Equivalent arguments apply to abiotic nodes (e.g. chemical species within nutrient stocks).

Each node is ascribed a single state variable, $n$, typically biomass or concentration, and the state of each node is locally stable around an equilibrium value, $k$ (the carrying capacity or physical equilibrium). All node state values are normalized to an arbitrary empirically-defined baseline condition, and the baseline state therefore has the dimensionless value $n = 1$. Each node equilibrium state ($k$) is directly affected by external forcings (imposed exogenous conditions, such as water temperature or pH), and it is the external forcings that are used to drive the model. Nodes within the system are connected with links carrying influences between the node states. It follows, from the stability of individual nodes, that a system-level equilibrium condition exists. Each node is part of an ecosystem that is intrinsically stable under 'background' conditions. This is the starting point for much theoretical work (e.g. Feng and Bailey 2018), and from an empirical perspective, is evidenced by abundant records of ecosystem stability during stable climatic regimes of, for example, the Holocene epoch (Willis and others 2010). Under these stable conditions, there would be no lasting <u>net</u> influences between interacting nodes, and this is carried in to the model framework: the influence of a node ('1') on another node ('2') is zero if node '1' is in its equilibrium state. When away from equilibrium, the state of a node has a tendency to return to the equilibrium state, and while away from equilibrium it exerts some influence on nodes to which it is connected. For example, if node '1' (predator) is increased, a negative influence is exerted on node '2' (prey); if '1' is exactly at its equilibrium, it exerts no influence on '2'. Even though in this case there implicitly continues to be predation of '1' on '2', there is no <u>net</u> effect on '2' at the population level (averaged over time), and so the influence of '1' on '2' is zero. Consequently, it is not strictly necessary to include all elements of an ecosystem, as there is no explicit energy balance to be achieved. External factors can effect both the equilibrium ($k$) and the node state ($n$) directly, and these interactions are specific to each node. In the following section, a more detailed description is provided.

3.0     Model description

3.1     Node state dynamics

The dynamics of the node states are determined by three components: (i) $R_j$, the rate of relaxation towards a local equilibrium (the local equilibrium being either $k$ or zero, described below); (ii) changes in local equilibrium due to external forcing; (iii) $I_j$, the combined direct influences of other nodes and by external forcing ($j$ is the node index). Each node state evolves according to

$$f(n) = \frac{dn}{dt} = R(n) + n\Gamma \qquad \text{(Eq.1a)}$$

A schematic indicating interactions between nodes, and the influence of external forcing, is shown in Figure 1, which serves as a reference for the descriptions provided in the following sections. The rate of change of $n$ is subject to a constraint on its upper value, the maximum (biologically set) growth rate, $r_{max}$. We therefore re-write Eq.1a as

$$f(n) = \frac{dn}{dt} = min(R(n) + n\Gamma, nr_{max}) \qquad \text{(Eq.1b)}$$

The time evolution of $n$ is calculated as

$$n(t) = \int_0^t f(n)\, dt \qquad \text{(Eq.2)}$$

where $t$ is time. Results presented in this paper are all numerical solutions (implemented in MATLAB® version R2017b (9.3.0.713579), using code written by RB).

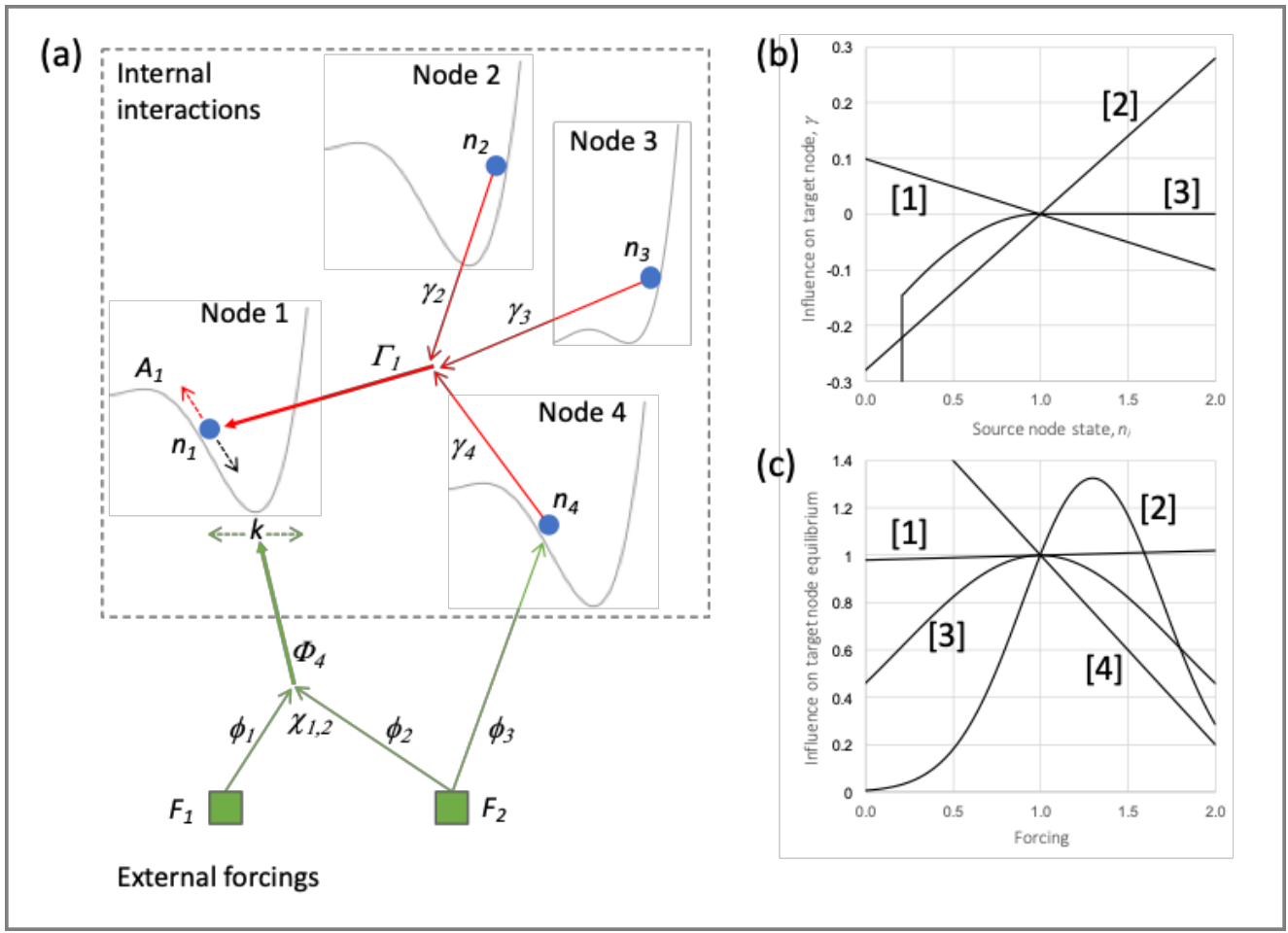

Figure 1 (a) A schematic representation of node interactions and influences by external forcings. States $n_2, n_3, n_4$ (filled blue circles) are away from their respective equilibria, and are exerting individual influences $\gamma_i$ (which combine to $\Gamma_i$) on the state of node 1, $n_1$. The dashed red arrow indicates the influence exerted (in this case) by the other nodes, $\Gamma_i$, and the black dashed arrow the attraction to the equilibrium, $k$. Also shown for node 1 is $A_1$, indicating the position of an unstable equilibrium, below which the state is attracted to $n = 0$ (see also Figure 2). External forcings $F_1$ and $F_2$ both affect the equilibrium ($k$) of node 1; their individual influences ($\phi_i$) are combined (under the influence of the multiplicative coefficient $\chi$) to $\Phi_4$, which affects change of $k$. External forcing $F_2$ also directly affects the state of node 4, $n_4$. See main text for further description. (b) Examples of inter-node interactions, where [1] indicates a negative influence of the source node on the target node (e.g. a predator's influence on a prey), [2] indicates a positive influence of the source node on the target node (e.g. a prey's influence on a predator), and [3] represents a more complicated response, with a saturating response at high node-state values, and a threshold response at low values. An example might be an abiotic node representing oxygen concentration, where there exists an absolute maximum benefit limit, a non-linear dependence below this limit, where biotic performance is impeded, and a threshold value below which the influence is so large (and negative) that it overwhelms all other influences. (c) Examples of forcing interactions, where relationships between forcing and equilibrium state include examples of weak positive [1], Gaussian ([2] and [3]), and strong negative [4].

3.2     Relaxation to the local equilibrium

As described above, there are two types of nodes in this model, biotic and abiotic. In both cases, an equilibrium ($k$) exists, towards which the node state is attracted. This 'relaxation' of the node state towards its equilibrium $k$ is depicted in Figure 2(a), which shows a single potential well (a 'basin of attraction') and example time series (Figure 2(b)).

The rate of relaxation of the node state towards its local (node-specific) equilibrium is defined by one of three functions. For the first type, relatively simple biotic nodes, the dynamics are dictated by a logistic model (Eq.3a), which yields an asymmetric potential and characteristic 'S-shaped' asymptotic growth. In this case, each node has a single stable state at $n_j = k_j$ (the 'carrying capacity'). In the absence of other influences, $n_j \to k$ as $t \to \infty$.

For some biological populations, a stable viable equilibrium is only possible above some critical population level (the 'strong Allee effect'; Courchamp and others 2008). Below this level, populations experience positive feedback on inefficiencies that further reduce the population (increased difficulties finding mates, or genetic bottlenecks are examples). This is not expected to be the case for all organisms, and debate persists over the existence and strength of Allee effects for marine organisms (e.g. Gregory and others 2010; Hutchings 2014). An option to impose Allee effects is retained and is achieved by specifying a lower unstable boundary state at $n_j = A_j$ (Fig. 1(a)). In this case, in the absence of other influences, $n_j \to 0$ when $n_j < A_j$, and $n_j \to k_j$ when $n_j > A_j$ (Eq.3(b)) (Figure 2(c), 2(d)).

Lastly, for abiotic nodes, the relaxation rate is proportional to the deviation from equilibrium $k_j$, yielding a symmetrical potential, and exponential recovery following perturbation (Eq.3c).

$$R_j = \begin{cases} n_j r_{0_j} \left(1 - \dfrac{n_j}{k_j(\Phi)}\right) & \text{(Eq.3a)} \\[2ex] n_j r_{0_j} \left(1 - \dfrac{n_j}{k_j(\Phi)}\right)\left(\dfrac{n_j}{A_j} - 1\right) & \text{(Eq.3b)} \\[2ex] r_{0_j}\left(k_j(\Phi) - n_j\right) & \text{(Eq.3c)} \end{cases}$$

where $r_{0_j}$ is the intrinsic recovery/growth rate. The equilibrium stable state $k_j$ is a function of $\Phi$, which combines the effects of multiple external forcings, and is defined in §3.3.

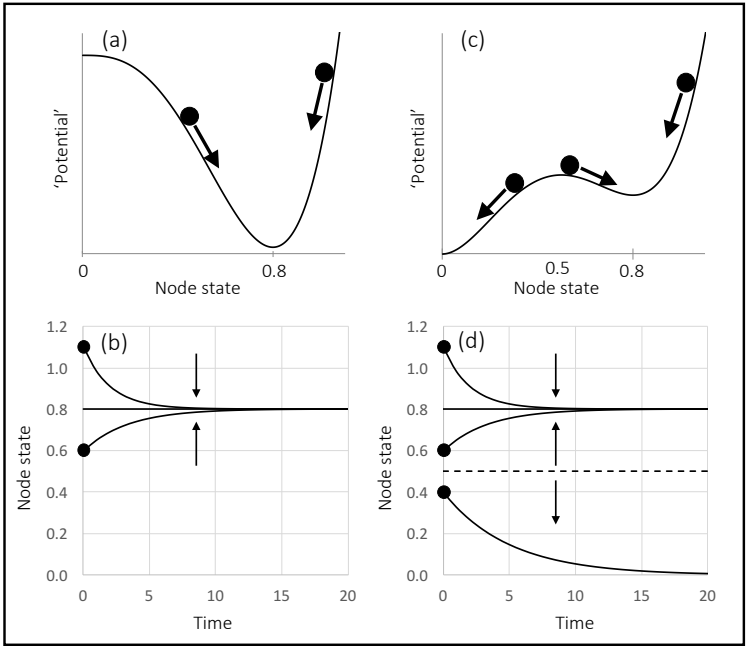

Figure 2 The relaxation of a node state to its local equilibrium. (a) The potential well indicates the rate of relaxation towards equilibrium at 0.8. (b) Examples, relevant to (a), showing the state changing over time. (c) The presence of an unstable equilibrium (e.g. an Allee effect) at $n = 0.5$ is shown. (d) Time series of node states relevant to (c), where the unstable equilibrium is indicated (at 0.5) by the dashed line.

## 3.3 External forcing

There are $N_F$ external (exogenous) forcing factors and it is changes in these forcings over time that principally drives the model. Changes in external forcing are specific to each node and can have two types of effects: direct effects on node states (which are treated the same way as internal interaction effects, in increasing/reducing the node state value; e.g. fishing pressure), and effects on the node equilibrium (e.g. temperature, nutrient concentration). Changes in the equilibrium $k$ ultimately drive changes in the state of the node ($n$) indirectly (with some lag) because, in the absence of other influences, $n_i \rightarrow k_i$ over time (when $n_i > A_i$, if applicable).

### 3.3.1 Response to individual forcings

The form of the response of $k_i$ to changes in each relevant forcing is defined by a function specific to each node/forcing pair, $\varphi_{m,j}$ (forcing $m$, node $j$) (note the node equilibrium value ($k$) is normalized by the equilibrium state under baseline forcing, meaning $k_j = 1$ under baseline forcing conditions (see §3.3.2 and §3.3.3)) In the case of biotic nodes, $\varphi_{m,j}$ represents a tolerance curve, while for abiotic nodes, $\varphi_{m,j}$ represents the relevant physical processes (e.g. the dependence of dissolved $O_2$ on temperature). The choice of function for $\varphi_{m,j}$ is arbitrary, and informed by data/theory. Here we provide example cases of polynomial and Gaussian functions (Eq.4a,b) used in the model implementation described in §4-6.

$$\varphi_{m,j}(F_m) = \begin{cases} \sum_{q=0}^{n_q} a_q \, F_m^q & \text{(Eq.4a)} \\ a(b.\exp(-(F_m-b)^2/(2c^2))) + d & \text{(Eq.4b)} \end{cases}$$

where $a, b, c, d$ are fitting parameters chosen such that the function passes through the point (1,1) (see Figure 1(c)).

A peak-shaped function for the response of $k_i$ (e.g. Figure 1(c)[2] and [3]) is a natural choice in some cases (e.g. temperature responses), as it reflects well-known tolerance curves described for many species (e.g. Pörtner 2009); alternatively, monotonic functions may be used to describe dependencies over relatively limited forcing ranges (e.g. pH responses; Busch and McElhany 2016). The baseline forcing condition does not necessarily coincide with the maximum (optimal) state of any node (compare lines [2] and [3] in Figure 1(c)), or the system as a whole, and therefore a move away from baseline forcing can be beneficial or detrimental, as required. In an equivalent way, abiotic nodes also have an equilibrium condition (for example, if a dissolved oxygen state is briefly perturbed from equilibrium, such as through changes in biological activity, it should in time return to this equilibrium). Changes in external forcing (e.g. water temperature) can also therefore affect the equilibrium conditions of abiotic nodes in the model.

As change in external forcing affects change in node equilibrium, the resilience of biotic nodes (the shape of their potential well) is also modified. Under benign ('low stress') conditions (leading to increase in equilibrium/carrying capacity), the potential of the node becomes steeper; Figure 3). Under favourable conditions, therefore, both post-perturbation recovery rates of node states and their resistance to perturbations is greater (higher resilience). The converse is also true, and biotic nodes have flatter potential wells, so lower recovery rates, and greater sensitivity to perturbations, under more stressful forcing conditions. If an Allee threshold is present, then changes in forcing can drive the equilibrium either towards or away from this unstable point. Reducing the difference between the stable and (fixed) unstable state causes a flattening of the basin (compare nodes 3 and 4 in Figure 1), to the extent that it eventually becomes a downward slope towards zero, i.e. a full loss of resilience. One consequence is that biotic node states become more susceptible to variability (noise) in the forcing as overall stress increases. Additionally, node states can be 'dragged' towards the unstable boundary more easily through the interactions of other nodes, if the node is in this less resilient condition. External forcing can create conditions for the collapse of nodes and this effect can be transmitted through interactions with other nodes to the wider network, potentially leading to cascades or the appearance of alternative stable system states (e.g. §4.2.4).

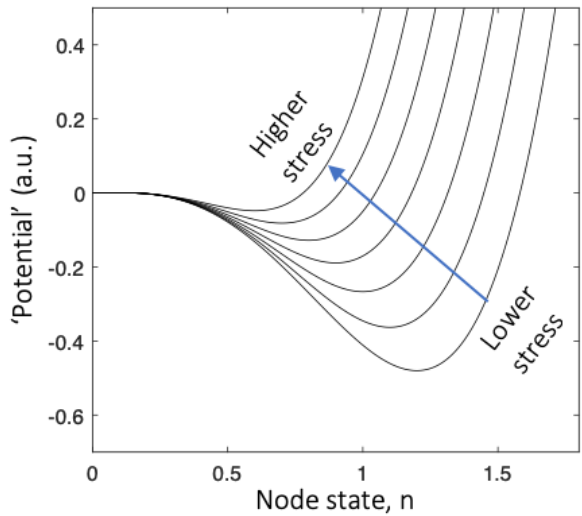

**Figure 3** As the equilibrium ($k$) is reduced (under stress from external forcing), the shape of the potential well is affected, here flattening as the equilibrium ($k$) approaches the unstable equilibrium at $n = 0.1$. This is associated with a reduction in the resilience of the node state, as discussed in §3.3.1).

3.3.2   Variation in forcings over time

The model can in principle be parameterized for time steps of any length (those that adequately capture relevant processes). It may be necessary to account for the variance in forcing that occurs within the chosen time resolution, particularly if nodes are disproportionately sensitive to extremes in forcing. To parameterize within-time step variation in forcing, each external forcing comprises a time series for each of the parameters necessary to define a chosen distribution ($G(F)$) of forcing values at each model timestep. In the present case we use a generalized normal distribution, and the three time series specify parameters for the central value (mean, $\mu_F$), scale (standard deviation, $\sigma_F$), and shape (to allow for skewness, $\kappa_F$). A time-dependent distribution for each forcing variable can therefore be defined. These distributions are multiplied by relevant response functions ('tolerance curves', for biotic nodes) specific to each node (see below), to provide an 'effective forcing' effect on each node over time. This makes the model sensitive to temporal variation in both the 'mean' forcing and the precise form of the distribution.

To include inter-time-step variability in forcing, additive white noise (scaled appropriately) is added independently to each of the forcing parameters appropriately (see §3.2.4).

The 'effective forcing' for each year (per forcing) is then the integral of the product of the forcing distribution, $G(F)$, and the forcing response function, $\varphi(F)$, defining $\alpha^*$,

$$\alpha^* = \int_0^\infty \varphi(F)G(F)dF \qquad \text{(Eq.5)}$$

(which is calculated numerically in the present case). Naturally, if the shape, width or position of the forcing distribution changes over time, then so too does $\alpha^*$ (an implicit assumption here is that statistical variation in the ordering of the fine scale variations in forcing is not important, and averaged effectively over successive forcing time steps.). There is a normalization of $\alpha^*$ by $\alpha_0^*$, the value of $\alpha^*$ under baseline conditions.

$$\alpha = \alpha^*/\alpha_0^* \qquad \text{(Eq.6)}$$

The value $\alpha$ therefore defines the equilibrium value ($k$) resulting from a single forcing.

3.3.3    Response to multiple forcings

In the presence of a single forcing factor, an equilibrium node state $k_j$ would be defined solely by the relevant value of $\alpha$. Under the influence of multiple forcings, incremental contributions from each source (towards the equilibrium values dictated by the relevant functions) must be combined. The form of this combination of influences is illustrated in Eq.7 (and specific details are outlined below). In this simplified example, three influences $x_1, x_2, x_3$ are combined to a single influence $y$.

$$y = (a_1 x_1 + a_2 x_2 + a_3 x_3) + (b_1 x_1 x_2 + b_2 x_1 x_3 + b_3 x_2 x_3) \qquad \text{(Eq.7)}$$

In the first bracket, coefficients $a_i$ scale the addition of the single effects, and these are set to 1 in all present cases. In the second bracket, coefficients $b_i$ provide similar scaling for all pairwise multiplicative components, and can be chosen to determine the nature of the combined effect ($b_i > 1$ for synergistic, $b_i < 1$ for antagonistic, $b_i = 0$ to remove the multiplicative contribution). Inclusion of only pairwise multiplicative combinations is a simplification (as compared to including all possible combinations of effects). In principle there is no barrier to including additional combinations, but the realities of finding data to justify parameter choices for real-world model applications is typically prohibitive. We also implicitly assume the coefficients are single-valued, and not functions of other stressors.

Following the form of Eq.7, the combined effect of multiple external forcings on any given node, including their pairwise multiplicative interactions, is described by an integrating function, $\Phi$. First, we define an $N_F \times N_F$ lower triangular matrix, $\chi^{Fk}$, which stores relevant coefficients: diagonal elements $(\chi_{i,j=i})$ are the coefficients for additive terms and off-diagonal elements $(\chi_{i,j \neq i})$ the multiplicative terms (*sensu* Eq.7). Then,

$$X_{j,t+dt} = k_{j,t} + \sum_{m \in [1:N_F]} \left(\chi^{Fk}_{m,m}(\alpha_{0_m} - \alpha_{1_m})\right) +$$

$$\sum_{m \in [1:N_F]} \sum_{c \in [1:m-1]} \left(\chi^{Fk}_{m,c}\left(\frac{\alpha_{1_m}}{\alpha_{0_m}} \cdot \frac{\alpha_{1_c}}{\alpha_{0_c}} - 1\right)\right) \qquad (Eq.8)$$

where $k_{j,t}$ is the equilibrium state value of node $j$ at time $t$, and $\alpha_{0_m}$ and $\alpha_{1_m}$ follow from Eq.6,7 above (subscripts 0 and 1 referring to values of $\alpha$ at time $t$ and $t + dt$ respectively. The second term sums over all $N_F$ forcing factors ($\chi^{Fk}_{j,m} = 0$ where forcing $m$ does not affect node $j$), and calculates the contribution from additive effects. The third term sums over all $((N_F^2 - N_F)/2)$ relevant multiplicative pairings of forcings (in the example above, Eq.7, $N_F = 3$ and 3 such pairings exist).

Irrespective of the combined forcings, the equilibrium state of each node has a feasible lower limit of zero, and this is enforced by

$$\Phi_j = max\{X_j, 0\} \qquad (Eq.9)$$

### 3.3.4 A note on variability (noise) in the forcing

A distinction is made between true variation in node state and measurement noise. True variation must be physically and/or biologically possible, as for example, populations cannot be expected to increase faster than their biological limit under ideal conditions. For this reason, the dynamics are deterministic, and noise is included independently on each of the forcing terms rather than on the node states directly. For simulation of empirical measurement data, statistical (measurement) noise could be added to the model output, with an appropriate choice of model, but this is not included in the present version. An example of adding Gaussian noise to the forcing time series is given in §5.

3.4        Direct influences on node state

Each node is connected to at least one other node with a directed link (Figure 1(a)), and an influence is exerted from a source node to a target node whenever the source node is not at equilibrium. The nature of the influence depends on the specific relationship between the nodes (e.g. predation, facilitation), and is determined by both the interaction function specific to each connection (Figure 1(b)) and the states of the sources and target nodes. Interactions have the potential therefore to drive increases or decreases in the target node state value. When multiple nodes exert influences on a single target node (e.g. combinations of predator and prey nodes), those influences are combined additively, and applied to the target node (Figure 1(a)). While internal (inter-node) influences drive changes between connected nodes, these influences are not fully prescriptive of the target node state. They provide a 'pressure' that drives the node state to increase or decrease, and the precise response of the target node is determined both by this net 'pressure' and by the local gradient of its potential (Figures 1,2). There is convergence to a meta-stable condition where the (incoming) inter-node influences are balanced by the tendency to return to equilibrium (this is discussed further in §4.2.4).

External forcing can also provide a pressure directly on the node state, as indicated in Figure 1(a), where 'Forcing 3' directly affects the state of 'Node 4'. As the effect of pressure (from one node to another, or from direct external forcing) depends on the gradient of the potential (the 'basin shape') of the target node, the same level of influence can potentially affect different nodes in different ways, and can affect the same node differently at different times.

Whether direct effects on node states are due to internal interactions (between nodes) or external forcing, the effects are treated in the same way in the model. The influence on the state of node $j$ is $\gamma_{x,j}$ (index $x$ representing either $i$ or $m$, for node state and forcing values respectively), by an arbitrary function informed by data/theory. Typically, the function passes through point $(1,0)$, meaning that when the influencing state/forcing is equal to 1, the baseline equilibrium condition, no influence is exerted (see Figure 1(b)). In some cases it is appropriate to have a $(0,0)$ point of reference for external influences (zero influence when the forcing value is equal to zero). For example, if the baseline condition for pollution is zero, this would need to be reflected in the choice of function representing its influence.

Functions are specified for each individual interaction, and the majority of these would be expected to be predator/prey relationships. Increases in predator biomass exerts a negative influence on prey ($\gamma < 0$), while increasing prey biomass exerts a positive influence on its predators ($\gamma > 0$). In the simplest cases, these relationships are linear, but non-linear relationships including those with discontinuities may be necessary (see Figure 1(b) and caption). In addition to the functional form of the node-node interactions, the interaction strength ($\gamma$) between predators and prey can be scaled in proportion to prey availability, where multiple prey nodes are present, to account for 'prey switching' behaviour of predators (a choice to favour more abundant prey), and this is shown below.

Equation 10 defines $\gamma$ in the case of an inter-node effect, with options for polynomial or Gaussian functions (used in the example simulations described below; §4-7). This equation also holds for external influences on node state (substituting forcing values for the node state value ($n_i$) and setting $\rho_{j,i} = 0$).

$$\gamma_{i,j}(n_i) = \begin{cases} \rho_{j,i} + \sum_{q=0}^{n_q} a_q\, n_i^q & \text{(Eq.10a)} \\ \rho_{j,i} + a_{i,j} + \left[\left(b_{i,j} exp\left(-(n_i - c_{i,j})^2 / (2 d_{i,j}^2)\right)\right)\right] & \text{(Eq.10b)} \end{cases}$$

where $a, b, c, d, q$ are fitting parameters, and $\rho_{j,i}$ adjusts the interaction coefficient for the effects of prey switching.

$$\rho_j = \beta\left(\frac{\gamma_k}{\gamma_0} - 1\right), \qquad \text{(Eq.11)}$$

where $\beta \in [0,1]$ controls the proportional effect of prey switching (0 for no prey switching, 1 for the predation interaction scaled wholly in proportion to prey availability, and intermediate values for switching fractional proportions of the predation effect);

$$\gamma_k = \left(\sum_{k=1}^{N_p} \gamma_{0,j}\right) \gamma_j^* \Big/ \sum_{k=1}^{N_p} \gamma_k^* \qquad \text{(Eq.12)}$$

and

$$\gamma_j^* = \gamma_{0,j} - \gamma_{0,j}\left(\frac{n_j}{\sum_{k=1}^{N_p} n_k} - 1\right) \qquad \text{(Eq.13)}$$

$N_p$ is the number of prey nodes of node j.

Values of $\gamma_{x,j}$ ($x$ being either node $i$ or forcing $m$) represent the sole effect of $x$ on node state $n_j$, and when multiple influences are present for any node, these are combined additively as $\Gamma_j$ (see Eq.14 & 15). The incremental change in the rate of change of $n_j$, is $\gamma_{x,j}^1 - \gamma_{x,j}^0$, where $\gamma_{x,j}^0 = \gamma_{x,j}(t)$ and $\gamma_{x,j}^1 = \gamma_{x,j}(t+dt)$. For each node, values of $\gamma_{x,j}^0$ and $\gamma_{x,j}^1$ are generated for each connected node and external forcing (zero in the case of unconnected nodes or non-interacting external factors), and four vectors hold these values: (i) $\zeta^{n(0)}$ and $\zeta^{n(1)}$ both of length $N_n$, for inter-node interactions, and $\zeta^{F(0)}$ and $\zeta^{F(1)}$ both of length $N_F$, for external interactions. To simplify notation these vectors are combined as $\zeta^0 = [\zeta^{n(0)}, \zeta^{F(0)}]$ and $\zeta^1 = [\zeta^{n(1)}, \zeta^{F(1)}]$, where $[\cdot,\cdot]$ denotes concatenation. In addition, $\zeta^a$ is a vector of length $N_n + N_F$ holding associated weights, included here for generality, but set to unity in the present case. The combined direct effects of multiple external forcings and multiple inter-node interactions are combined additively for each node (dropping index $j$) as

$$\Gamma = \Gamma_t + \sum_{p \in [1:N_n+N_F]} \left( \zeta_p^a (\zeta_p^1 - \zeta_p^0) \right) \qquad \text{(Eq.14)}$$

Interactions from other nodes, and from external forcings that directly affect node states, are then

$$I_j = \Gamma_j n_j \qquad \text{(Eq.15)}$$

### 3.5 Model parameter uncertainty considerations

In the current formulation of the model, each node requires at least four prescriptive parameters to account for its behaviour and response to forcing, plus a multiplicative coefficient ($\chi$) for each pairwise forcing combination (affecting each node), plus at least one parameter per interaction with other nodes. In any given model instantiation, each of these parameters is uncertain to some degree, and there must be a facility within the model implementation to incorporate this uncertainty in to probabilistic results. The model is sufficiently computationally light that statistical sampling within random distributions of each parameter is feasible, and in this case the model produces distributions of node states over time, from which probabilistic interpretations can be made. As the number of parameters is relatively high (of the order of hundreds), we use Latin Hypercube Sampling to efficiently samples the space, with each dimension of the hypercube representing each uncertain parameter. Examples are discussed in the following sections.

4.0     Model behaviour

We can make a number of immediate observations based on the model description outlined above, and in the following sections these observations will be explored using a model representing a simplified marine ecosystem. We first describe briefly the model, then evaluate of some of its basic properties, then explore impacts of changes in forcing, effects of variability in forcing, and the effect of synergistic interactions between stressors.

4.1 A simplified marine ecosystem model

This model represents a generic temperate marine ecosystem, with 16 nodes for the common major functional groups (plus a detritus node). The structure of the ecosystem model is shown graphically in Figure 4, where two-way interactions, principally predator-prey, are indicated by lines connecting nodes. The model includes representations of the plankton community, several larger fish functional groups, and apex predators (toothed whales, pinnipeds and sharks). A separation between pelagic and benthic communities is made in both the benthic fauna (representing the major invertebrate groups) and the benthopelagic fish. A (bentho-pelagic) coupling between these communities is included through prey-predator interactions.

The model is specified at an annual resolution, and for simplicity the between-node interaction functions (Eq.10(a), $q = 1$) are linear. Model parameters for these interactions were estimated from relevant example dietary composition, consumption-biomass ratios, assimilation efficiency and handling time (full description in Table A1, Appendix A).

Linear trends in temperature, pH, pollution concentration (heavy metals, persistent organic pollutants) and fishing pressure (on relevant nodes) are used to drive the model over a 50-year period (see Table A2, Appendix A for node equilibrium responses). Three different scenarios are used, and these are based on IPCC scenarios for temperature and pH, and arbitrary estimates for fishing pressure and pollution (see Table 1). In some model runs, Gaussian noise is added (independently) to each forcing (see §5.1). Node-specific responses to each forcings affect either the node's equilibrium (in the case of pH, temperature, and pollution) or its state (in the case of fishing pressure, here directly targeting 'large benthic fauna', 'small fish', 'bentho-pelagic fish', 'large fish' and 'squid'). Parameters defining these responses were based on experimental studies of survival rates, biological performance and stock assessment data. The response of each node to temperature forcing follows a Gaussian form, and all others are linear. While it is possible to incorporate in this model the indirect effects of forcings, such as bio-accumulation of pollution or bycatch effects of fishing, these were not included in the present version.

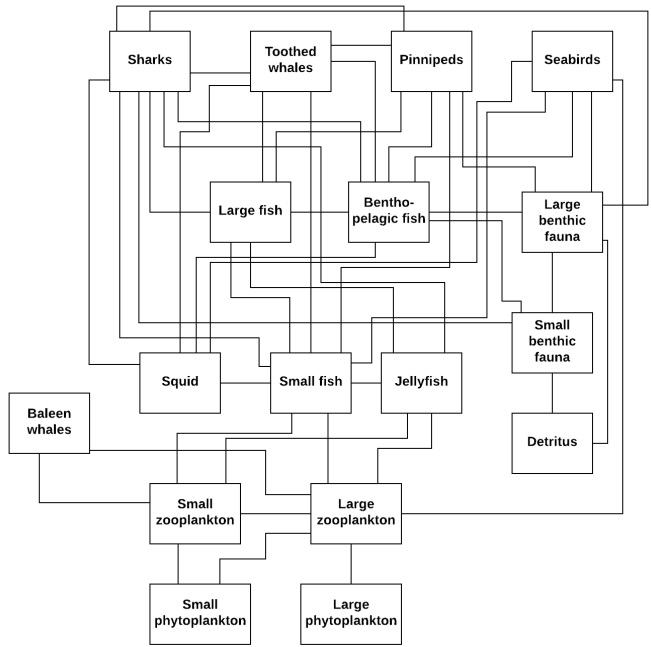

**Figure 4** Nodes and interactions in the simplified marine ecosystem model described in the main text. Each biotic node represents an aggregation of commonly-occurring marine functional groups, and there is also included a detritus node. Lines represent two-way connections between nodes (principally predator/prey interactions).

4.2 Rudimentary observations

As an initial assessment of the model's behaviour, we can make a number of observations from the equations presented in §3. These give insights in to the model, and in some cases provide minimal 'reality checks'. Relevant numerical solutions based on the simplified ecosystem model are also provided, and discussed further in §5).

4.2.1     Nodes remain stable at zero

It should be the case that future biomass changes depend in part on the current biomass levels, and that if biomass drops to zero, this state is then indefinitely maintained (no spontaneous regeneration). Reference to Eq.1 confirms $n$ appears as a multiplier in all terms, hence $dn/dt = 0$, if $n = 0$, as required (see also Appendix B, Figure B1).

### 4.2.2 Constant equilibrium observed with external forcing at background levels

When external forcings are at baseline conditions, Eq.6 yields $\alpha = 1$, which defines the node equilibrium as $k = 1$. As time progresses, the summation terms in Eq.8 remain at zero, and $k$ remains constant. With all nodes at their stable equilibrium, $k$, there are no net influences exerted between nodes, hence all states remain constant over time at unity (Appendix B, Figure B2). This demonstrates the nature of the interactions in this model as being the net interactions, and only non-zero when node states are away from their equilibrium states.

### 4.2.3 Unsupported biotic nodes go extinct

Where trophic interactions to a node represent its complete dietary intake, setting those prey nodes to zero should result in its extinction. In the model this happens because interactions from prey nodes become increasingly negative as prey biomass is reduced. A necessary condition for the relevant extinction is that (when prey nodes are set to zero) the sum of negative influences is greater than the sum of positive influences, such as those from any relevant predator release. That is, the lack of predation on a node cannot make up for a lack of prey; likewise, beneficial forcing conditions cannot provide an escape from zero prey. The sum of negative influences must therefore be sufficient to counter the tendency to return to $k$ under such conditions ($I < 0$ and $|I| > |R|$ when all relevant prey nodes are at zero). Examples are provided in Appendix B for the parameters chosen for the simplified ecosystem described, where it is seen that nodes do go to zero when prey is removed (Appendix B, Figure B3).

### 4.2.4 Alternative (quasi-)stable states (where $R = -I$)

The state of any node ($n_j$) will over time tend towards a 'quasi-stable' equilibrium condition where the three main influences on the dynamics (direct influences from other nodes, direct influences from external forcing, and relaxation towards the local equilibrium) are in balance, i.e. where $R = I$. That is, we expect the state of each node to approach a balance, whereby the effects of 'pushes' from external influences ($I$) are balanced by the 'pull' from the tendency to return to equilibrium ($R$). For example, if a node was removed, new equilibria would be reached across the ecosystem (see example in Figure 5(a). Alternatively, adding a fishing pressure would reduce targeted node biomass levels to a point of balanced between removal rate and regrowth rate ($r$), and other indirectly affected nodes would reach new equilibria due to these changes in biomass (see examples in Figure 5(b)).

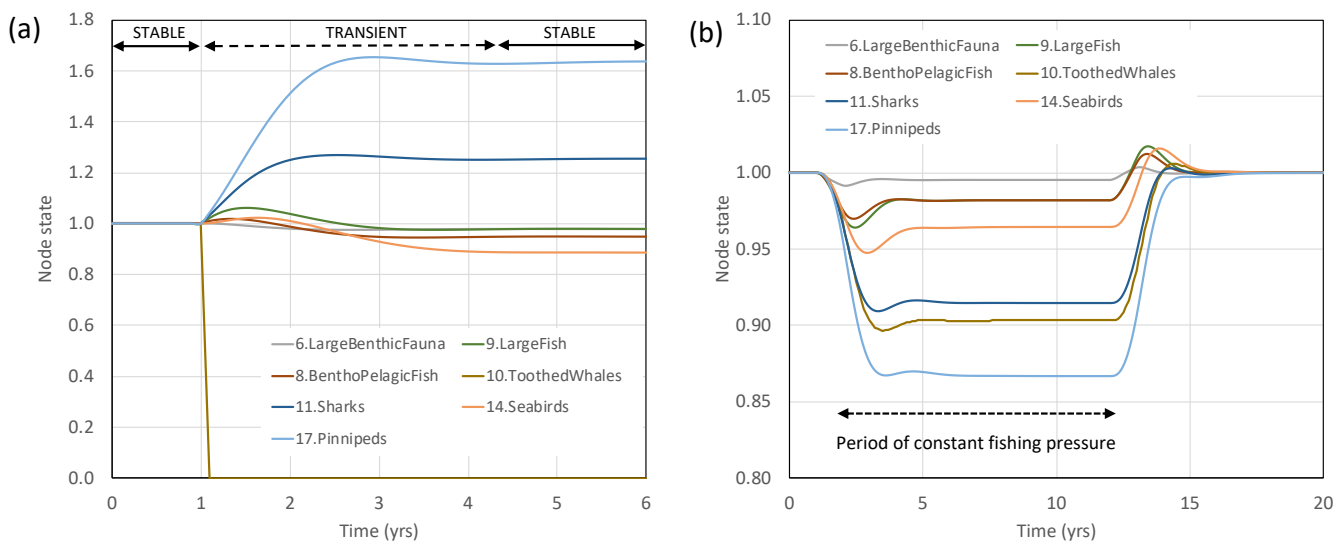

**Figure 5** Establishment of alternative stable states. Showing results for example nodes: (a) The removal of 'toothed whales' biomass (node 10) results in establishment of a new stable state for the ecosystem. The initial stable conditions (under background forcing) are indicated, followed by the transient response to the biomass removal, and the establishment of a new stable condition after year 4 of the simulation. (b) Application of strong fishing pressure (directly affecting nodes: 'large benthic fauna', 'small fish', 'bentho-pelagic fish', 'large fish' and 'squid') in year 2-12 results in a novel (quasi-)stable state. The progression of [stable → transient (onset of fishing) → quasi-stable → transient (fishing halted) → stable] can be observed.

4.2.5    Effects of external forcing on node resilience

External forcing not only changes the carrying capacity ($k$) of biotic nodes, but also the shape of the potential well around the stable equilibrium point (see Figure 3, Eq. 3a,b). All else being equal, if a node is at its equilibrium state $k$, a small perturbation of that state ($dn \ll k$) will be followed by a recovery to $k$. Reference to Eq.2 shows that where an imposed perturbation results in a small fractional reduction in $n$ (and ignoring inter-node influences and influences from the unstable equilibrium at $A$), the initial relaxation rate (and therefore the relaxation time) is directly proportional to $k$. Hence, under more stressful forcing conditions, there is a reduction in both the node carrying capacity ($k$) and in the post-perturbation recovery rate. In addition, because the magnitude of $R$ (in the vicinity of $k$, when $n \neq k$) is reduced with $k$, the effect of a given (negative) perturbation is also increased, as the quasi-equilibrium reached (§4.2.3), when $R = -I$, occurs at a lower value of $n$. Thus, the same perturbation can have different effects on the same node under different levels of ambient stress. If the node has an unstable state (the Allee effect – see §3.2), greater levels of stress will mean successively smaller perturbations are required to force $n_j < A_j$, hence tipping the node in to a potentially collapsing transition to $n = 0$. Under higher levels of externally imposed stress, the node therefore has a lower carrying capacity, a slower recovery post-perturbation, and an increased sensitivity to shocks. Such changes in node response (particularly if multiple nodes are affected by the external stressor) are likely to reduce the resilience at the level of the whole system (this is explored further below).

4.3     Perturbation experiments

In these experiments we use isolated single perturbations of individual node states as a way to probe model behaviour, and specifically the nature of internal interactions and the associated dynamics. The main purpose is to assess whether model behaviour is as expected, in terms of the sign of changes in associated nodes, their rates and magnitudes of change.

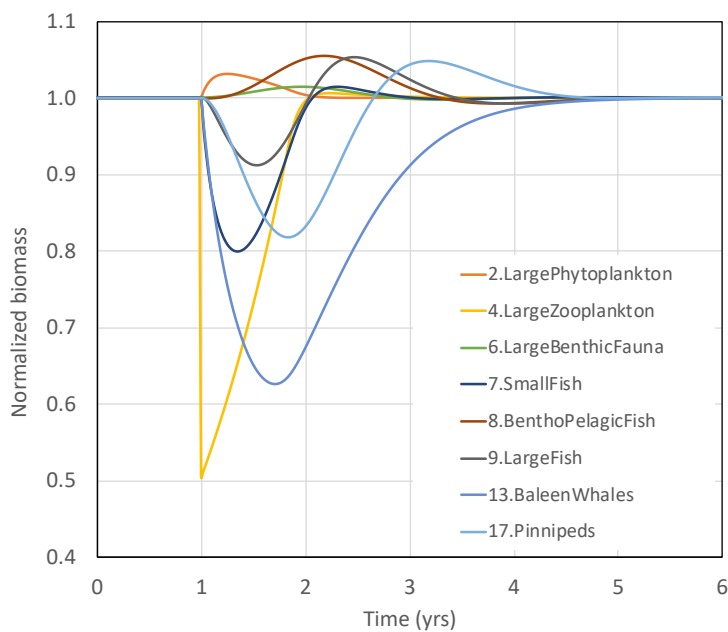

**Figure 6** Selected results from a perturbation experiment. Under constant background-level forcing, the perturbation imposed was an instantaneous reduction of the 'Large Zooplankton' biomass (to 50%) at the end of year 1. Subsequent changes in other nodes represent first-, second- and third-order (food web) effects, and these are discussed in the main text.

Example results are shown in Figure 6, where the perturbation is to the node representing large zooplankton (LZP), and is an instantaneous reduction to 50% at year one of the simulation. First-order responses occur in nodes directly linked to LZP: large phytoplankton biomass increases, due to predatory release, whilst nodes predating on the LZP (small fish and jellyfish) decline immediately due to a loss of prey (albeit in different amounts, due to differences in their nodes characteristics). Second-order effects (in nodes two links apart from LZP) are seen, for example, in declining 'large fish' (which initially suffer from the loss of 'small fish' prey); a third-order effect is the initial increases in bentho-pelagic fish, due to predator release from 'large fish', which then provides impetus for a recovery of the 'large fish'. Examples of many such effects can be seen in Figure 6 (informed by Figure 4). The details in each case depend on node position within the ecosystem, dietary composition, prey-switching and growth rates. In this example perturbation, all node states have settled back into their original equilibrium after ten years (i.e. the system maintains resilience following small perturbations, and the ensuing changes are reversible). We find no evidence that the model is not behaving in line with expectations (see commentaries, and for additional examples, in Appendix B, Figures B4 & B5), and this provides a minimal level of validation for such a simplified model. We note that in some cases, when the perturbation is relatively strong, the initial equilibrium condition is not recovered and a novel stable configuration is achieved (with different relative biomass distribution).

5.0     Model experiments: forcing strength, synergy strength and forcing variability

In this section, we describe experiments designed to explore the effects of forcing strength, strength of synergies, and noise in the forcings, using the simplified ecosystem model described in §4.1. These experiments are forced by two sources of stress common, in the context of global change, to many marine ecosystems: rising water temperature and reductions in pH; plus two typically more localized stressors: increases in pollution concentration, and fishing pressure. As described in §3, forcings have effects that are individual to each node (either on carrying capacity, e.g. temperature) or on node state (e.g. fishing pressure), and not all nodes are affected by each forcing (see Table A2 of Appendix A for details). For simplicity, we hold the (within-timestep) forcing distribution shape constant and vary only the mean ($\mu_F$) over time (see §3.3.2). Model simulations run for 50 years, with a linear change to the mean value of the forcing, $\mu_F$ (see Table 1). The slopes of these changes, under three different forcing scenarios, 'mild-', 'moderate-' and 'intense-forcing' are given in Table 1 (see caption for further details). The nature of the response of each node to these forcings is defined by parameters $a, b, c, d$ (Eq. 4), and the specific choice of function (in the present case either linear or Gaussian), and these details are provided in Table A2 (Appendix A).

To assess the effect of synergy strength between forcings, we take a simplified approach of assigning all multiplicative coefficients ($\chi$) a common value, and repeating simulations with different common values. We use three scenarios: 'low-' ($\chi = 0$), 'medium-' ($\chi = 0.3$), and 'high-' ($\chi = 0.6$). These values were chosen following the work of Crain et al. (2008), who provide data on the sign and magnitude of interactive (pairwise) effects for 171 marine species (their Hedges-d statistic was re-formulated to provide estimates of the multiplicative coefficient, $\chi$, with the aid of additional data provided by Dr C. Crain (pers.com.). In forthcoming work, we present a more detailed ecosystem model, with $\chi$ values directly informed by empirical data, but for the present case we note that 0, 0.3 and 0.6 represent appropriate values for low, medium and high levels of synergy between stressors.

As stresses increase on biotic nodes, and the equilibrium (carry capacity) is reduced, there is a flattening of the potential of each node (see §3.3.1 and Figure 3), and an inevitable increase in the sensitivity to direct influences (internal or external) on the node state. As random variations (noise) in external forcings are a source of such perturbations, it is expected that the strength of individual forcings, and degree of synergy between the forcings, would have a significant effect on the sensitivity of the ecosystem to noise. To assess the impact of noise in the forcing, we add independent Gaussian noise to each forcing time series, such that $\mu_F \sim N(\hat{\mu}_F, \sigma_F)$. To provide a common baseline for scaling the magnitude of the noise for each of the forcings, we take the change in forcing over the 50 year simulation period for the 'moderate' scenario (Table 1), and multiply that value by a 'noise coefficient' (either 0, 0.5, or 1) to provide the value for $\sigma_F$. Example results for the model experiments outlined here are provided in the following section (§6).

|         | Temperature (K) |            | pH        |            | Pollution conc. |            | Fishing pressure |            |
|---------|-----------------|------------|-----------|------------|-----------------|------------|------------------|------------|
| Scenario | Absolute (50 yrs) | Relative (per yr) | Absolute (50 yrs) | Relative (per yr) | Absolute (50 yrs) | Relative (per yr) | Absolute (50 yrs) | Relative (per yr) |
| Background | 0 | 0 | 0 | 0 | 0 | 0 | 0 | 0 |
| 'Mild'   | 1.0 | 0.000070 | -0.14 | -0.000342 | 0.10 | 0.002 | 0.05 | 0.0010 |
| 'Moderate' | 2.0 | 0.000140 | -0.17 | -0.000420 | 0.20 | 0.004 | 0.10 | 0.0020 |
| 'Intense' | 2.9 | 0.000205 | -0.20 | -0.000499 | 0.30 | 0.006 | 0.15 | 0.0030 |

**Table 1** Forcing scenarios. Values for each external forcing indicate the absolute change in forcing (here $\mu_F$; see §5.0; temperature, K; pH, pH units; pollution and fishing pressure are relative to an assumed baseline) over the 50 year simulation period, and also shown as the annual relative rate of change. Forcing values can be expressed as normalized values (relative to baseline conditions) or in absolute units, so long as the response/tolerance curves are expressed in those same units.

6.0     Results

6.1     Forcing and synergy strength

Example model output timeseries (for selected nodes) are shown in Figure 7 for the cases of mild-forcing/low-synergies, moderate-forcing/medium-synergies, and intense-forcing/high-synergies. Figure 8 provides a summary for each of the nine combinations of forcing and synergy conditions, where vertical bars indicate the difference in node state between year 0 and year 50 of the simulation period (blue for net loss of biomass, red for net gains). In this plot, the nodes are ordered roughly by trophic level, and this highlights the tendency of higher trophic nodes to respond more strongly to the severity of forcing and increased synergies between forcings. It is also noteworthy that phytoplankton show an increase in biomass, due to reduced predation. Similarly, the response of several nodes is not monotonic over time, due to changes in the balance of predator/prey biomass over time (as the stress from external forcing increases).

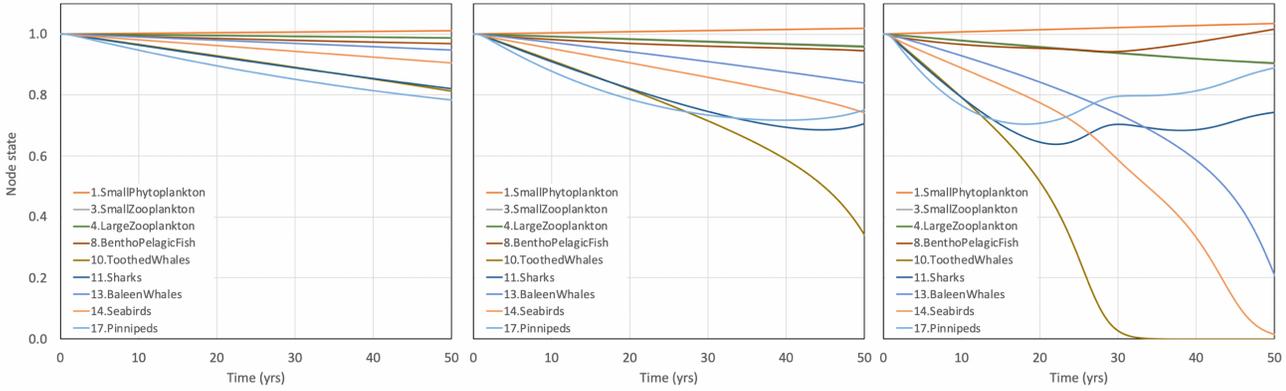

**Figure 7** Example time series for example nodes under linear changes in external forcing. (a) 'mild' forcing and $\chi = 0$; (b) 'moderate forcing' and $\chi = 0.3$; (c) 'intense forcing' and $\chi = 0.6$. A full explanation of the model experiment conditions is provided in the main text. Figure 8 provides a summary of full results and additional time series results are shown in Appendix B, Figure B6.

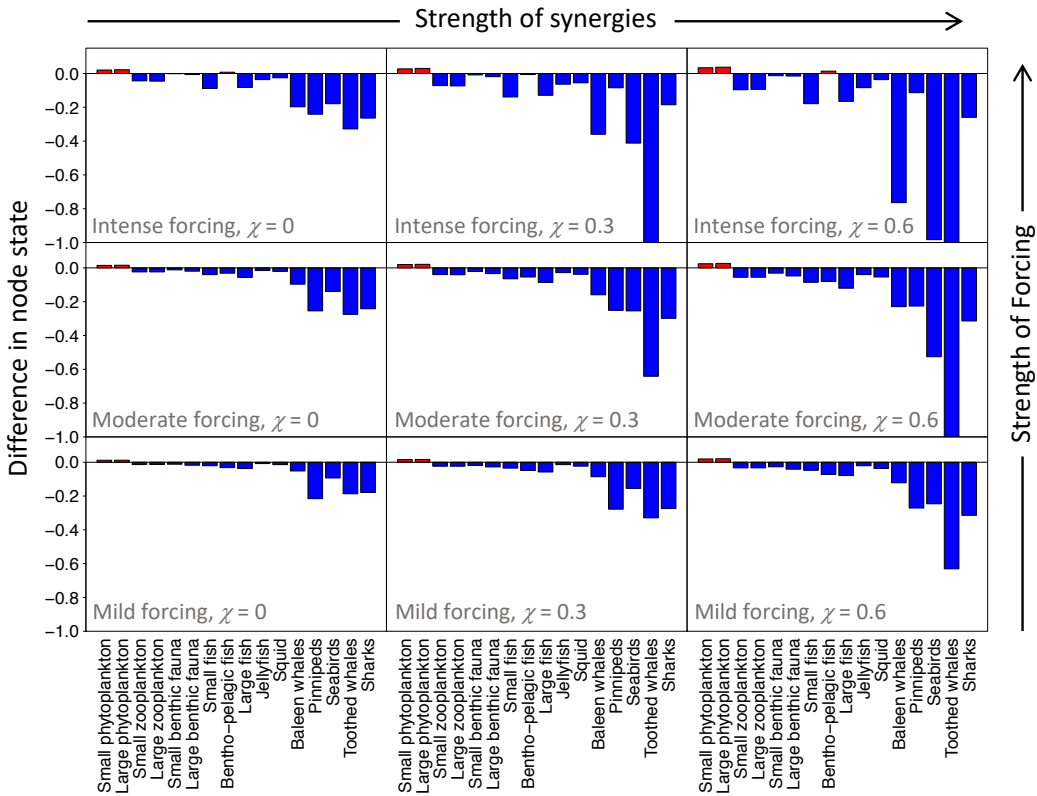

**Figure 8** The combined effects on biomass of forcing strength and strength of synergies between forcings. Forcing strength scenarios are defined in Table 1, and the strength of the synergies ($\chi$, see §3.3.3) is indicated in the figure. Blue bars represent net loss, and red bars net gain of biomass over the 50 year simulation period.

## 6.2 Effects of noise

Noise in forcing produces associated noise in the results, and in some cases has an effect on the nature of the model response to changes in forcing. Figure 9 shows example results for nodes 9 ('large fish') and 11 ('sharks'), where three different levels of noise have been added to the forcings (see previous section and figure caption for details); the figures show the distribution of final node states (at model year 50) following 1000 independent simulations. Nodes which remain resilient and at relatively high biomass (e.g. node 9), shows an approximately symmetrical broadening of the final state results similar in form to the Gaussian noise applied to the forcing (Figure 9(a)). For nodes which decrease in biomass significantly over the 50 year simulation (e.g. node 11 [sharks]), the effect of noise is more pronounced. In the case of node 11, there is also an asymmetry towards lower values, and a significantly increased probability of collapse to zero (Figure 9(b)).

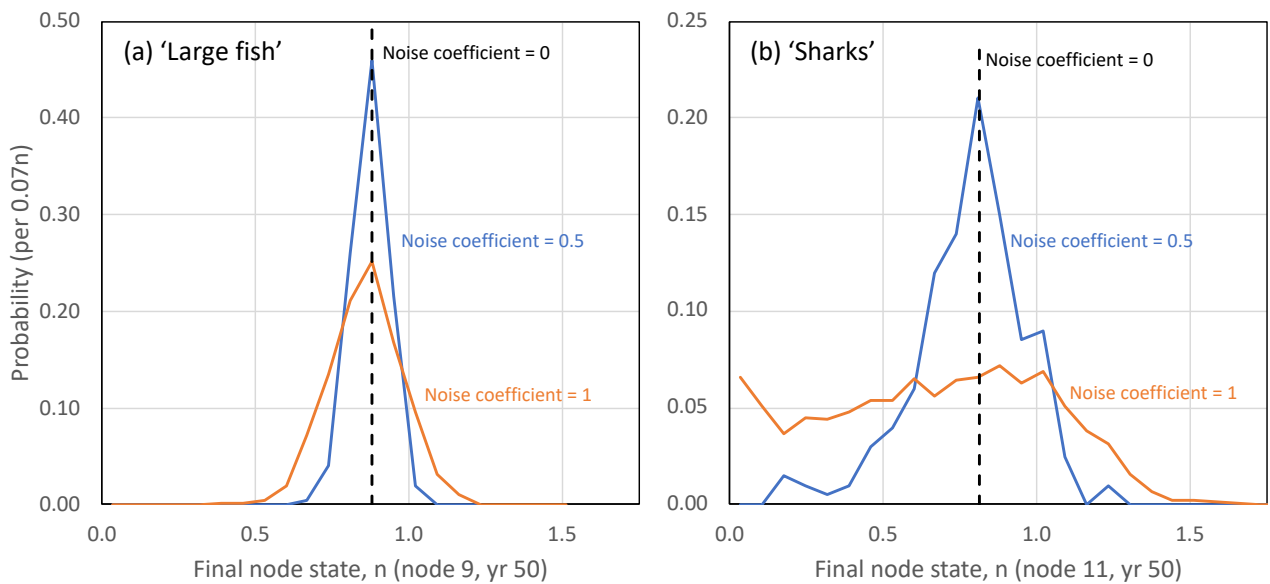

**Figure 9** The effect of noise in forcing on the final state of nodes 9 ('large fish') and 11 ('sharks'). Simulations were run for 50 yrs under the 'moderate' forcing scenario, with $\chi = 0.3$. Larger noise coefficients mean greater random variability in the forcing time series, as described in §5.0.

## 6.3 Effects of parameter uncertainty

To provide a sense of the implications of parameter uncertainty on model output, Figure 10 shows example results in which 148 model parameters were assigned a symmetrical uniformly-distributed uncertainty, with a range equal to 10% of the central value (additional details in caption). Latin hyper-cube sampling (§3.5) was used to re-sample each parameter within its bounds, and model variant outputs were generated for each parameter set. Figure 10 shows results for 100 such re-samplings, and in each case the model output using the central values is indicated by the dashed black lines (the thinner continuous lines are the variants).

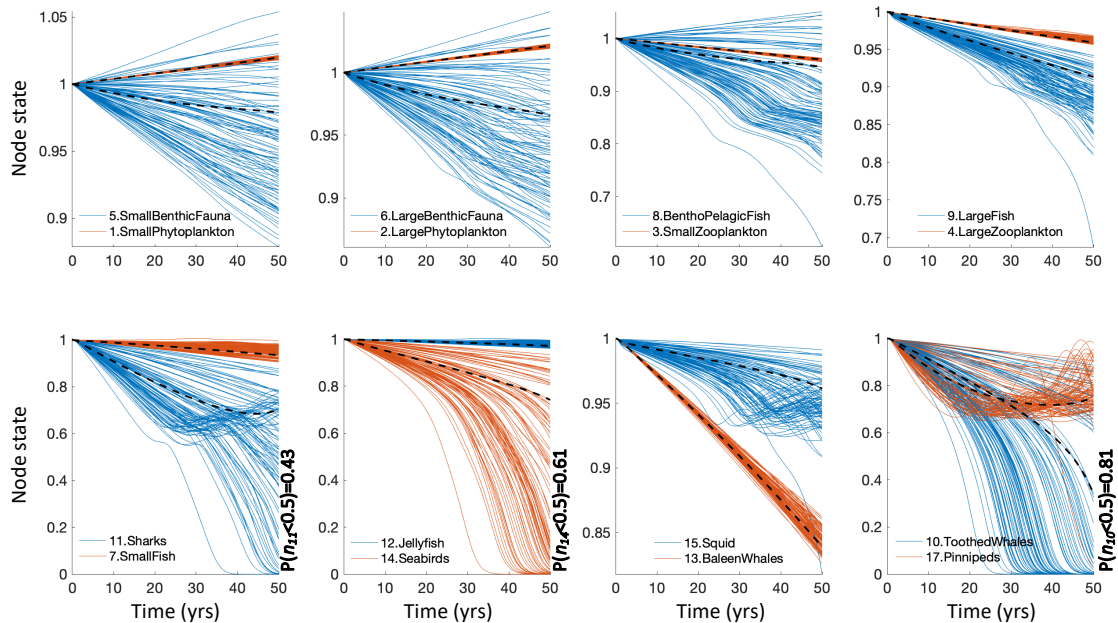

**Figure 10** Repeated model runs with statistical sampling of key model parameters. All simulations were run under the 'moderate' forcing scenario (with zero noise) and $\chi = 0.3$. Parameter distributions were Uniform, with central values given in Table A1 & A2 of Appendix A, and total width of 10% of the central value. 148 parameters were re-sampled, accounting for all inter-node interactions, and all responses to external forcings). Along the lower panel, and for the more significantly affected nodes, the probability of node value ($n$) being below 0.5 at year 50 is shown (43% for sharks, 61% for seabirds, 81% for toothed whales).

## 7.0 Discussion

The problem being tackled at the heart of this work is the difficulty in assessing the impact of multiple simultaneous stressors on ecosystems generally, and on marine ecosystems specifically. The effect on organisms of changes in individual conditions (e.g. pH) are well characterised in some cases, but the effects of interactions between many such changes (e.g. water temperature, pH, pollution, habitat degradation), in the context of wider ecosystem interactions (e.g. associated changes in interspecific competition) are seldom known. There is need then for a framework within which the possible effects of interactions between multiple stressors can be examined both for individual species (or functional groups) and on the ecosystem as a whole. OSIRIS is an attempt to provide such a framework.

In all ecosystem models there exists varying levels of certainty about the structural nature of the ecosystem (what interacts with what) and parameters chosen to describe those interactions. For this reason, the OSIRIS model framework is highly flexible in terms of defining the ecosystem structure, the nature of interactions between the ecosystem elements (nodes), and the nature and impacts of individual external forcing and their interactions (at the level of individual nodes). The parameter $\chi$ provides means to control the sign and magnitude of the multiplicative interaction between forcings (i.e. producing antagonistic, purely additive or synergistic effects) for each node. This flexibility allows the influence of design choices to be evaluated relatively easily. The ability to produce probabilistic outputs (through the use of latin hypercube sampling, as shown in Figure 10) provides an efficient means of propagating parameter uncertainty through to the final results, and is a strength of the current model and our aim to keep the model sufficiently computationally light to facilitate this re-sampling approach.

Given the current level of uncertainty in quantifying ecosystem relationships (specifically the degree of antagonism/synergy that exists between sets of stressors for individual species or functional groups), we do not make claims that OSIRIS (or even perhaps other models) can be predictive of future ecosystem states. Rather, we imagine using this approach to assess scenarios and assess sensitivities. For example, from a baseline of knowledge of effects of individual impacts, an important question is that of how strong synergistic relationships between stressors must be in order for expected forcings to have significantly negative ecosystem effects (however these are defined). Then, the question of whether such levels of synergy are within uncertainty bounds of current estimates, or feasible on ecological/biological/physiological grounds is one that can be asked of data or future research programmes.

The implementation of the model presented in this paper is highly simplified. For example, it has relatively few functional groups represented, coarse (annual) time resolution, it is non-spatial, there is no age structure in the biological nodes, and the same value of $\chi$ is used for all nodes. These simplifications, which clearly reduce the model's realism, were chosen for this particular implementation, and are not characteristics of the OSIRIS model itself. Other implements (in preparation for publication) include considerably finer time resolution, nodes assigned to life stages, data-informed estimates of $\chi$ per functional group, and other refinements. Nonetheless, the present model implementation is informed by relevant data (Appendix A), and is expected to be a reasonable representation of the general behaviour of temperate marine ecosystems, providing example model outputs and examples of high-level insights possible with the current approach.

The transient responses to perturbations (Figures 5 and 6) indicate the dynamic responses produced by this model implementation are not obviously incorrect; for example, the first- and higher-order responses observed in Figure 6, and the magnitudes and rates of these changes (see also Appendix B). The model is able to produce irreversible changes in stable states (e.g. Figure 5a), and also reversible quasi-stable states under constant perturbation/forcing (e.g. fishing pressure, as shown in Figure 5b). The possibility of ecosystems being forced in to novel stable/quasi-stable states has received much attention in recent years (e.g. Scheffer et al., 2001), and it is important the model is capable of capturing these dynamics. The possibility of predicting such transitions from observational data remains a significant challenge (ref.), but even relatively coarse models have a part to play in uncovering identifiable precursors (whether these be time series properties or structural changes; e.g. Bailey 2010; Scheffer et al., 2012).

Model experiments described in §5 investigated the effects of forcing strength, strength of synergies, and variability (noise) in the forcing. To constrain the experimental setup, temperature and pH forcing were based on IPCC projections (Bopp et al. 2013; IPCC 2013; Doney et al. 2014) and the range of $\chi$ values (from 0 to 0.6) used for synergies between forcings were based on equivalent values derived from the work of Crain et al. (2008) (the range of 0.6 is approximately the standard deviation of the equivalent empirical data). Increasing either the severity of forcing (columns of Figure 8), or the strength of synergies between forcings (rows of Figure 8), results in greater decline of biomass over the 50 year simulation period; increasing both (diagonal of Figure 8; see also Figure 7) results in a considerably enhanced effect, and more so at higher trophic levels. This experimental setup is most likely overestimating true effects, as all nodes are given the same level of synergy between all forcings, whereas the evidence strongly suggests a high degree of heterogeneity, and antagonistic interactions between forcings for some species. However, that the rather extreme response occurs for values spanning only 1 standard deviation of the empirical data is noteworthy. Higher trophic levels appear to be most sensitive to these effects, and a non-linear response is observed in relevant time series (e.g. Figure 7).

Greater levels of stress (from stronger forcing) cause a flattening of the potential well of biotic nodes (Figure 3 and §3.3.1), and as the synergy between forcings is increased, the strength of forcing becomes a more crucial factor in determining the resilience of individual nodes. Variability (noise) in forcing has the effect of broadening the results, as shown in distributions of final node states (at year 50) in Figure 9, and these effects are stronger at higher trophic levels. For individual nodes, there is an asymmetry in the shape of the potential which means reductions in biomass due to (symmetrical) noise are expected to be longer lasting than the equivalent benefits (see Figure 2 and 3; note the post-perturbation recovery to equilibrium in Figure 2 is slower when the perturbation reduces the biomass compared to when biomass is increased by the perturbation, even though the magnitude of the perturbation is smaller). This asymmetry is seen in Figure 9b at higher noise levels, and importantly, causes a collapse of the biomass in a disproportionately large fraction of the simulations. It is possible that variability in climate conditions will increase under global climate change, and if synergy between forcings is towards the higher end of current estimates, may be a significantly more important impact than changes in mean conditions.

It is important in all analytical/modelling work that the effects of uncertainty are correctly handled, and that uncertainties (that exist in all parameters in all models) are not ignored, and are propagated through to the final results. Within the framework of a precautionary approach, the widening of model outputs serves as a reminder of the dangers of false precision in model (and data analysis) outputs. Explicit inclusion of parameter uncertainty is a strength of the present model, and results shown in Figure 10 (and see caption) provide an illustrative example of the raw model output (time series) and the kinds of probabilistic conclusions that naturally follow.

Modelling ecosystems necessitates a coarsening and simplification across various domains: spatial, temporal and biological/ecological. As discussed above, the current model implementation lacks representation of space, and implicitly aggregates spatially-resolved inter-node interactions and localized forcings. The model framework allows for the splitting of existing nodes in to 'sub-nodes' in different locations. This would provide means of simulating, for example, spatial diffusion and localized forcings. The model in the present implementation runs at annual time resolution, and increasing this to finer timesteps may be advantageous, if nodes respond to variation at these timescales in ways that cannot be aggregated over a year. However, this raises the difficulty in finding adequately resolved data to parameterize nodes at these finer scales. The present model version aggregates species in to functional groups, and then represents those groups by a single state variable (biomass). The framework allows for hierarchical interactive structures 'within' nodes to be represented (these are simply other nodes, and the network structure reflects the hierarchy). For example, what would be otherwise aggregated 'species' nodes, could be represented by multiple sub-nodes reflecting specific properties such as age, relevant life history traits, sex or combinations of these. Likewise, there is no limit to the number of stressors that can be included, or the level of complexity of the interaction terms, and these can be linked to any of the sub-nodes with any functional response. What is presented here represents a 'bare bones' approach designed to show proof of concept, and to allow for relatively transparent interpretation. Adding greater complication will in principle allow fuller quantitative model/data comparisons, and ultimately a finer level of interpretation, but comes with a cost of greater data needs and more complex outputs.

## 6.0     Conclusion

The OSIRIS model provides a framework to explore the currently uncertain effects of multiple stressors on marine ecosystems. The model is highly flexible, and is shown to be capable of simulating dynamic responses in line with expectations for such ecosystems. For the relatively simple temperate marine ecosystem model presented, the effect on biomass of synergistic interactions between the effects of individual forcings is found to be substantial. Within reasonable estimates for the forcings and the degree of potential synergy between those forcings, the model system shows significant effects at the higher trophic levels, and a strong susceptibility to variability (noise in the forcings). The model can be adapted to different ecosystems relatively easily, and is sufficiently computationally light that statistical re-sampling within the uncertainty bounds of individual parameters is possible, meaning the outputs can be probabilistic, and the quoted level of uncertainty in the model output is not underestimated. There are many potential uses, such as assessing sensitivities to specific local or global forcing scenarios, or the effects on confidence levels of uncertainties in individual parameters (and hence the value of additional research to reduce such uncertainty). This work also underlines the significant potential risk incurred in treating stressors on ecosystems as individual and additive, particularly in light of the large uncertainty in the degree of synergy that exists when many stressors are present, both for individual species and ecosystems as a whole.

# References

IPCC. 2013. Climate Change 2013: The Physical Science Basis. Contribution of Working Group I to the Fifth Assessment Report of the Intergovernmental Panel on Climate Change. In: Stocker TF, Qin D, Plattner G-K, Tignor M, Allen SK, Boschung J, Nauels A, Xia Y, Bex V, Midgley PM (eds.)). Cambridge University Press, Cambridge, United Kingdom and New York, NY, USA, 1535 pp.

Kroeker KJ, Kordas RL, Crim R, Hendriks IE, Ramajos L, Singh GS, Duarte CM, Gattuso J-P. 2013. Impacts of ocean acidification on marine organisms: quantifying sensitivities and interaction with warming. Global Change Biology, 19: 1884-1896.

Marshall KN, Kaplan IC, Hodgson EE, Hermann A, Busch DS, McElhany P, Essington TE, Harvey CJ, Fulton EA. 2017. Risks of ocean acidification in the California Current food web and fisheries: ecosystem model projections. Global Change Biology, 23: 1525-1539.

Parker LM, Ross PM, O'Connor WA, Pörtner H-O, Scanes E., Wright JM. 2013. Predicting the response of molluscs to the impact of ocean acidification. Biology, 2: 651-692.

Piggott JJ, Townsend CR, Matthaei CD. 2015. Reconceptualizing synergism and antagonism among multiple stressors. Ecology and Evolution, 5: 1538-1547.

Polovina JJ. 1984. Model of a coral reef ecosystem I. The ECOPATH model and its application to French Frigate Shoals. Coral Reefs, 3: 1-11.

Pörtner H-O. 2009. Oxygen- and capacity-limitation of thermal tolerance: a matrix for integrating climate-related stressor effects in marine ecosystems. The Journal of Experimental Biology, 213: 881-893.

Pörtner H-O, Karl DM, Boyd PW, Cheung WWL, Lluch-Cota SE, Nojiri Y, Schmidt DN, Zavialov PO. 2014. Ocean systems. Climate Change 2014: Impacts, Adaptation, and Vulnerability. Part A: Global and Sectoral Aspects. Contribution of Working Group II to the Fifth Assessment Report of the Intergovernmental Panel on Climate Change. Field CB, Barros VR, Dokken DJ, Mach KJ, Mastrandrea MD, Bilir TE, Chatterjee M, Ebi KL, Estrada YO, Genova RC, Girma B, Kissel ES, Levy AN, MacCracken S, Mastrandrea PR, White LL, editors. Cambridge, United Kingdom and New York, NY, USA, Cambridge University Press: 411-484.

Rodolfo-Metalpa R, Houlbrèque F., Tambutté É, Boisson F., Baggini C., Patti FP, Jeffree R., Fine M., Foggo A., Gattuso J-P, Hall-Spencer JM. 2011. Coral and mollusc resistance to ocean acidification adversely affected by warming. Nature Climate Change, 1: 308-312.

Rogers AD, Laffoley Dd'A. 2013. Editorial, Introduction to the special issue: The global state of the ocean; interactions between stresses, impacts and some potential solutions. International Programme on the State of the Ocean 2011 and 2012 workshops. Marine pollution bulletin, 74: 491–494.
32

Acknowledgements

Funding for this work was provided by Ocean Conservancy and Stanford University Catalyst Programme. The authors are grateful to Dr. K. Crain and Dr Ben Halpern for providing data on stressor interactions, and to Andreas Merkl, Dr George Leonard and Dr Anna Zivian for valuable discussions during the model building phase.